\documentclass[10pt]{scrartcl}
\usepackage{hyperref}
\usepackage{epsfig}
\usepackage[numbers]{natbib}
\usepackage{amsmath}
\usepackage{amsbsy}
\usepackage{amsfonts}
\usepackage[latin1]{inputenc}
\usepackage{url}

\typearea{10}

\providecommand{\email}[1]{E-mail:\href{mailto:#1}{\texttt{#1}}}
\providecommand{\dprod}{\! \cdot \!}%
\providecommand{\wprod}{\! \wedge \!}
\begin{document}
%

\title{Euclidean formulation of general relativity}
\author{\textbf{J. B. Almeida}\\ \normalsize
{Universidade do Minho, Departamento de F\'isica,}\\\normalsize {4710-057
Braga, Portugal.}\\\normalsize \email{bda@fisica.uminho.pt}}

\pagestyle{myheadings} \markright{Euclidean formulation of general relativity
\hfill J. B. Almeida }


\date{\normalsize }

%
%

%
\maketitle

\begin{abstract}                
A variational principle is applied to 4D Euclidean space provided with a tensor
refractive index, defining what can be seen as 4-dimensional optics (4DO). The
geometry of such space is analysed, making no physical assumptions of any kind.
However, by assigning geometric entities to physical quantities the paper
allows physical predictions to be made. A mechanism is proposed for translation
between 4DO and GR, which involves the null subspace of 5D space with signature
$(-++++)$.

A tensor equation relating the refractive index to sources is established
geometrically and the sources tensor is shown to have close relationship to the
stress tensor of GR. This equation is solved for the special case of zero
sources but the solution that is found is only applicable to Newton mechanics
and is inadequate for such predictions as light bending and perihelium advance.
It is then argued that testing gravity in the physical world involves the use
of a test charge which is itself a source. Solving the new equation, with
consideration of the test particle's inertial mass, produces an exponential
refractive index where the Newtonian potential appears in exponent and provides
accurate predictions. Resorting to hyperspherical coordinates it becomes
possible to show that the Universe's expansion has a purely geometric
explanation without appeal to dark matter.
\end{abstract}
%

%
\section{Introduction}
According to general consensus any physics theory is based on a set of
principles upon which predictions are made using established mathematical
derivations; the validity of such theory depends on agreement between
predictions and observed physical reality. In that sense this paper does not
formulate a physical theory because it does not presume any physical
principles; for instance it does not assume speed of light constancy or
equivalence between frame acceleration and gravity. This is a paper about
geometry. All along the paper, in several occasions, a parallel is made with
the physical world by assigning a physical meaning to geometric entities and
this allows predictions to be made. However the validity of derivations and
overall consistency of the exposition is independent of prediction correctness.

The only postulates in this paper are of a geometrical nature and can be
condensed in the definition of the space we are going to work with:
4-dimensional space with Euclidean signature $(++++)$. For the sole purpose of
making transitions to spacetime we will also consider the null subspace of the
5-dimensional space with signature $(-++++)$. This choice of space does not
imply any assumption about its physical meaning up to the point where geometric
entities like coordinates and geodesics start being assigned to physical
quantities like distances and trajectories. Some of those assignments will be
made very early in the exposition and will be kept consistently until the end
in order to allow the reader some assessment of the proposed geometric model as
a tool for the prediction of physical phenomena. Mapping between geometry and
physics is facilitated if one chooses to work always with non-dimensional
quantities; this is easily done with a suitable choice for standards of the
fundamental units. In this work all problems of dimensional homogeneity are
avoided through the use of normalising factors for all units, listed in Table
\ref{t:factors}, defined with recourse to the fundamental constants: Planck
constant, gravitational constant, speed of light and proton charge.
\begin{table}[bt]
\caption{\label{t:factors}Normalising factors for non-dimensional units used in
the text; $\hbar \rightarrow$ Planck constant divided by $2 \pi$, $G
\rightarrow$ gravitational constant, $c \rightarrow$ speed of light and $e
\rightarrow$ proton charge.}
\begin{center}
\begin{tabular}{c|c|c|c}
Length & Time & Mass & Charge \\
\hline & & & \\

$\displaystyle \sqrt{\frac{G \hbar}{c^3}} $ & $\displaystyle \sqrt{\frac{G
\hbar}{c^5}} $  & $\displaystyle \sqrt{\frac{ \hbar c }{G}} $  & $e$
\end{tabular}
\end{center}
\end{table}
This normalisation defines a system of \emph{non-dimensional units} with
important consequences, namely: 1) all the fundamental constants, $\hbar$, $G$,
$c$, $e$, become unity; 2) a particle's Compton frequency, defined by $\nu =
mc^2/\hbar$, becomes equal to the particle's mass; 3) the frequent term
${GM}/({c^2 r})$ is simplified to ${M}/{r}$.

The particular space we chose to work with can have amazing structure,
providing countless parallels to the physical world; this paper is just a
limited introductory look at such structure and parallels. The exposition makes
full use of an extraordinary and little known mathematical tool called
geometric algebra (GA), a.k.a.\ Clifford algebra, which received an important
thrust with the introduction of geometric calculus by David Hestenes
\cite{Hestenes84}. A good introduction to GA can be found in \citet{Gull93} and
the following paragraphs use basically the notation and conventions therein. A
complete course on physical applications of GA can be downloaded from the
internet \cite{Lasenby99} with a more comprehensive version published recently
in book form \cite{Doran03} while an accessible presentation of mechanics in GA
formalism is provided by \citet{Hestenes03}.

\section{Introduction to geometric algebra}
We will use Greek characters for the indices that span 1 to 4 and Latin
characters for those that exclude the 4 value; in rare cases we will have to
use indices spanning 0 to 3 and these will be denoted with Greek characters
with an over bar. The geometric algebra of the hyperbolic 5-dimensional space
we want to consider $\mathcal{G}_{4,1}$ is generated by the frame of
orthonormal vectors $\{\mathrm{i},\sigma_\mu \}$, $\mu = 1 \ldots 4$, verifying
the relations
\begin{equation}
    \mathrm{i}^2  = -1,~~~~ \mathrm{i} \sigma_\mu + \sigma_\mu \mathrm{i}
    =0,~~~~
     \sigma_\mu
    \sigma_\nu + \sigma_\nu \sigma_\mu  = 2 \delta_{\mu \nu}.
\end{equation}
We will simplify the notation for basis vector products using multiple indices,
i.e.\ $\sigma_\mu \sigma_\nu \equiv \sigma_{\mu\nu}$. The algebra is
32-dimensional and is spanned by the basis
\begin{equation}
    \begin{array}{cccccc}
    1, & \{\mathrm{i},\sigma_\mu \}, &
    \{\mathrm{i} \sigma_\mu,\sigma_{\mu\nu} \}, &
    \{\mathrm{i} \sigma_{\mu\nu},\sigma_{\mu\nu\lambda}  \}, &
    \{\mathrm{i}I, \sigma_\mu I \}, & I; \\
    \mathrm{1~ scalar} & \mathrm{5~ vectors} & \mathrm{10~ bivectors} &
    \mathrm{10~ trivectors} & \mathrm{5~ tetravectors} & \mathrm{1~ pentavector}
    \end{array}
\end{equation}
where $I \equiv \mathrm{i}\sigma_1 \sigma_2 \sigma_3\sigma_4$ is also called
the pseudoscalar unit. Several elements of this basis square to unity:
\begin{equation}
 (\sigma_\mu)^2 = 1,~~~~ (i \sigma_\mu)^2=1,~~~~ (\mathrm{i}\sigma_{\mu\nu})^2 =1,
 ~~~~ (\mathrm{i}I)^2 =1;
\end{equation}
and the remaining square to $-1$:
\begin{equation}
    \mathrm{i}^2 = -1,~~~~(\sigma_{\mu\nu})^2 = -1,~~~~
    (\sigma_{\mu\nu\lambda})^2 = -1,~~~~(\sigma_\mu I)^2,~~~~I^2=-1.
\end{equation}
Note that the symbol $\mathrm{i}$ is used here to represent a vector with norm
$-1$ and must not be confused with the scalar imaginary, which we don't usually
need.

The geometric product of any two vectors $a = a^0 \mathrm{i} + a^\mu
\sigma_\mu$ and $b = b^0 \mathrm{i} + b^\nu \sigma_\nu$ can be decomposed into
a symmetric part, a scalar called the inner product, and an anti-symmetric
part, a bivector called the exterior product.
\begin{equation}
    ab = a \dprod b + a \wprod b,~~~~ ba = a \dprod b - a \wprod b.
\end{equation}
Reversing the definition one can write internal and exterior products as
\begin{equation}
    a \dprod b = \frac{1}{2}\, (ab + ba),~~~~ a \wprod b = \frac{1}{2}\, (ab -
    ba).
\end{equation}
When a vector is operated with a multivector the inner product reduces the
grade of each element by one unit and the outer product increases the grade by
one. There are two exceptions; when operated with a scalar the inner product
does not produce grade $-1$ but grade $1$ instead, and the outer product with a
pseudoscalar is disallowed.

\section{Displacement and velocity}
Any displacement in the 5-dimensional hyperbolic space can be defined by the
displacement vector
\begin{equation}
    \label{eq:displacement}
    \mathrm{d}s =\mathrm{i} \mathrm{d}x^0 + \sigma_\mu \mathrm{d}x^\mu;
\end{equation}
and the null space condition implies that $\mathrm{d}s$ has zero length
\begin{equation}
    \mathrm{d}s^2 = \mathrm{d}s \dprod \mathrm{d}s = 0;
\end{equation}
which is easily seen equivalent to either of the relations
\begin{equation}
    \label{eq:twospaces}
    (\mathrm{d}x^0)^2 = \sum (\mathrm{d}x^\mu)^2;~~~~
    (\mathrm{d}x^4)^2 = (\mathrm{d}x^0)^2 - \sum (\mathrm{d}x^j)^2.
\end{equation}
These equations define the metrics of two alternative spaces, one Euclidean the
other one Minkowskian, both equivalent to the null 5-dimensional subspace.

A path on null space does not have any affine parameter but we can use Eqs.\
(\ref{eq:twospaces}) to express 4 coordinates in terms of the fifth one. We
will frequently use the letter $t$ to refer to coordinate $x^0$ and the letter
$\tau$ for coordinate $x^4$; total derivatives with respect to $t$ will be
denoted by an over dot while total derivatives with respect to $\tau$ will be
denoted by a "check", as in $\check{F}$. Dividing both members of Eq.\
(\ref{eq:displacement}) by $\mathrm{d}t$ we get
\begin{equation}
    \label{eq:euclvelocity}
    \dot{s} = \mathrm{i} + \sigma_\mu \dot{x}^\mu = \mathrm{i} + v.
\end{equation}
This is the definition for the velocity vector $v$; it is important to stress
again that the velocity vector defined here is a geometrical entity which bears
for the moment no relation to physical velocity, be it relativistic or not. The
velocity has unit norm because $\dot{s}^2 =0$; evaluation of $v\dprod v$ yields
the relation
\begin{equation}
    \label{eq:vsquare}
    v \dprod v = \sum (\dot{x}^\mu)^2 = 1.
\end{equation}
The velocity vector can be obtained by a suitable rotation of any of the
$\sigma_\mu$ frame vectors, in particular it can always be expressed as a
rotation of the $\sigma_4$ vector.

At this point we are going to make a small detour for the first parallel with
physics. In the previous equation we replace $x^0$ by the greek letter $\tau$
and rewrite with $\dot{\tau}^2$ in the first member
\begin{equation}
    \label{eq:dtau2}
    \dot{\tau}^2 = 1 - \sum (\dot{x}^j)^2.
\end{equation}
The relation above is well known in special relativity, see for instance
\citet{Martin88}; see also \citet{Almeida02:2, Montanus01} for parallels
between special relativity and its Euclidean space
counterpart.\footnote{Montanus first proposed the Euclidean alternative to
relativity in 1991, nine years before the author started independent work along
the same lines.} We note that the operation performed between Eqs.\
(\ref{eq:vsquare}) and (\ref{eq:dtau2}) is a perfectly legitimate algebraic
operation since all the elements involved are pure numbers. Obviously we could
also divide both members of Eq.\ (\ref{eq:displacement}) by $\mathrm{d}\tau$,
which is then associated with relativistic proper time;
\begin{equation}
    \check{s} =  \mathrm{i}\check{x}^0 + \sigma_j \check{x}^j + \sigma_4.
\end{equation}
Squaring the second member and noting that it must be null we obtain
$(\check{x}^0)^2 - \sum (\check{x}^j)^2 = 1$. This means that we can relate the
vector $\mathrm{i}\check{x}^0 + \sigma_j \check{x}^j$ to relativistic
4-velocity, although the norm of this vector is symmetric to what is usual in
SR. The relativistic 4-velocity is more conveniently assigned to the 5D
bivector $\mathrm{i}\sigma_4\check{x}^0 + \sigma_{j4} \check{x}^j$, which has
the necessary properties. The method we have used to make the transition
between 4D Euclidean space and hyperbolic spacetime involved the transformation
of a 5D vector into scalar plus bivector through product with $\sigma_4$; this
method will later be extended to curved spaces.

Equation (\ref{eq:euclvelocity}) applies to flat space but can be generalised
for curved space; we do this in two steps. First of all we can include a scale
factor $(v = n \sigma_\mu \dot{x}^\mu)$, which can change from point to point
\begin{equation}
    \label{eq:refindex}
    \dot{s} = \mathrm{i} + n \sigma_\mu \dot{x}^\mu.
\end{equation}
In this way we are introducing the 4-dimensional analogue of a refractive
index, that can be seen as a generalisation of the 3-dimensional definition of
refractive index for an optical medium: the quotient between the speed of light
in vacuum and the speed of light in that medium. The scale factor $n$ used here
relates the norm of vector $\sigma_\mu \dot{x}^\mu$ to unity and so it deserves
the designation of 4-dimensional refractive index; we will drop the
"4-dimensional" qualification because the confusion with the 3-dimensional case
can always be resolved easily. The material presented in this paper is, in many
respects, a logical generalisation of optics to 4-dimensional space; so, even
if the paper is only about geometry, it becomes natural to designate this study
as 4-dimensional optics (4DO).

Full generalisation of Eq.\ (\ref{eq:euclvelocity}) implies the consideration
of a tensor refractive index, similar to the non-isotropic refractive index of
optical media
\begin{equation}
    \label{eq:vgeneral}
    \dot{s} = \mathrm{i} + {n^\mu}_\nu \dot{x}^\nu \sigma_\mu;
\end{equation}
the velocity is then generally defined by $v = {n^\mu}_\nu \dot{x}^\nu
\sigma_\mu$. The same expression can be used with any orthonormal frame,
including for instance spherical coordinates, but for the moment we will
restrict our attention to those cases where the frame does not rotate in a
displacement; this poses no restriction on the problems to be addressed but is
obviously inconvenient when symmetries are involved. Equation
(\ref{eq:vgeneral}) can be written with the velocity in the form $v = g_\nu
\dot{x}^\nu$ if we define the refractive index vectors
\begin{equation}
    \label{eq:gmu}
    g_\nu = {n^\mu}_\nu \sigma_\mu.
\end{equation}
The set of four $g_\mu$ vectors will be designated the \emph{refractive index
frame}. Obviously the velocity is still a unitary vector and we can express
this fact evaluating the internal product with itself and noting that the
second member in Eq.\ (\ref{eq:vgeneral}) has zero norm.
\begin{equation}
    \label{eq:vsquaregen}
    v \dprod v = {n^\alpha}_\mu \dot{x}^\mu {n^\beta}_\nu \dot{x}^\nu
    \delta_{\alpha \beta}=1.
\end{equation}
Using Eq.\ (\ref{eq:gmu}) we can rewrite the equation above as $g_\mu \dprod
g_\nu \dot{x}^\mu \dot{x}^\nu = 1$ and denoting by $g_{\mu \nu}$ the scalar
$g_\mu \dprod g_\nu$ the equation becomes
\begin{equation}
    g_{\mu \nu} \dot{x}^\mu \dot{x}^\nu =1.
\end{equation}
The generalised form of the displacement vector arises from multiplying Eq.\
(\ref{eq:vgeneral}) by $\mathrm{d}t$, using the definition (\ref{eq:gmu})
\begin{equation}
    \label{eq:dsgeneral}
    \mathrm{d}s = \mathrm{i}\mathrm{d}t + g_\mu \mathrm{d}x^\mu.
\end{equation}
This can be put in the form of a space metric by dotting with itself and noting
that the first member vanishes
\begin{equation}
    \label{eq:dt2general}
    (\mathrm{d}t)^2 = g_{\mu \nu} \mathrm{d}x^\mu \mathrm{d}x^\nu.
\end{equation}
Notice that the coordinates are still referred to the fixed frame vectors
$\sigma_\mu$ and not to the refractive index vectors $g_\mu$. In GR there is no
such distinction between two frames but \citet{Montanus01} clearly separates
the frame from tensor $g_{\mu\nu}$.

We are going to need the reciprocal frame \cite{Doran03}
$\{-\mathrm{i},g^\mu\}$ such that
\begin{equation}
    \label{eq:recframe}
    g^\mu \dprod g_\nu = {\delta^\mu}_\nu.
\end{equation}
From the definition it becomes obvious that $g_\mu g^\nu = g_\mu \dprod g^\nu +
g_\mu \wprod g^\nu$ is a pure bivector and so $g_\mu g^\nu = -g^\nu g_\mu$. We
now multiply Eq.\ (\ref{eq:dsgeneral}) on the right and on the left by $g^4$,
simultaneously replacing $x^4$ by $\tau$ to obtain
\begin{equation}
    \mathrm{d}s g^4 = \mathrm{i}g^4 \mathrm{d}t + g_jg^4
    \mathrm{d}x^j + \mathrm{d}\tau;~~~~g^4 \mathrm{d}s =
    g^4 \mathrm{i} \mathrm{d}t + g^4g_j
    \mathrm{d}x^j + \mathrm{d}\tau.
\end{equation}
When the internal product is performed between the two equations member to
member the first member vanishes and the second member produces the result
\begin{equation}
    \label{eq:transition1}
    (\mathrm{d}\tau)^2 = g^{44} \left[(\mathrm{d}t)^2 -g_{jk}\mathrm{d}x^j
    \mathrm{d}x^k \right].
\end{equation}
If the various $g_\mu$ are functions only of $x^j$ the equation is equivalent
to a metric definition in general relativity. We will examine the special case
when $g_\mu = n_\mu \sigma_\mu$; replacing in Eq.\ (\ref{eq:transition1})
\begin{equation}
    \label{eq:dtau2gen}
    (\mathrm{d}\tau)^2 = \frac{1}{(n_4)^2}\, (\mathrm{d}t)^2 -\sum\left(
    \frac{n_j}{n_4}\, \mathrm{d}x^j \right)^2.
\end{equation}
This equation covers a large number of situations in general relativity,
including the very important Schwarzschild's metric, as was shown in
\citet{Almeida04:1} and will be discussed below. Notice that Eq.\
(\ref{eq:dt2general}) has more information than Eq.\ (\ref{eq:transition1})
because the structure of $g_4$ is kept in the former, through the coefficients
$g_{\mu 4}$, but is mostly lost in the $g^{44}$ coefficient of the latter.
\section{The sources of space curvature}
Equations (\ref{eq:dt2general}) and (\ref{eq:transition1}) define two
alternative 4-dimensional spaces; in the former, 4DO, $t$ is an affine
parameter while in the latter, GR, it is $\tau$ that takes such role. The
geodesics of one space can be mapped one to one with those of the other and we
can choose to work on the space that best suits us.

The geodesics of 4DO space can be found by consideration of the Lagrangian
\begin{equation}
    L = \frac{g_{\mu \nu} \dot{x}^\mu \dot{x}^\nu}{2} = \frac{1}{2}\ .
\end{equation}
The justification for this choice of Lagrangian can be found in several
reference books but see for instance \citet{Martin88}. From the Lagrangian one
defines immediately the conjugate momenta
\begin{equation}
    v_\mu = \frac{\partial L}{\partial \dot{x}^\mu} = g_{\mu \nu} \dot{x}^\nu.
\end{equation}
The conjugate momenta are the components of the conjugate momentum vector $v =
g^\mu v_\mu$ and from Eq.\ (\ref{eq:recframe})
\begin{equation}
    v = g^\mu v_\mu = g^\mu g_{\mu \nu} \dot{x}^\nu = g_\nu \dot{x}^\nu.
\end{equation}
The conjugate momentum and velocity are the same but their components are
referred to the reverse and refractive index frames, respectively.

The geodesic equations can now be written in the form of Euler-Lagrange
equations
\begin{equation}
    \dot{v}_\mu = \partial_\mu L;
\end{equation}
these equations define those paths that minimise $t$ when displacements are
made with velocity given by Eq.\ (\ref{eq:vgeneral}). Considering the parallel
already made with general relativity we can safely say that geodesics of 4DO
spaces have a one to one correspondence to those of GR in the majority of
situations.

We are going to need geometric calculus which was  introduced by
\citet{Hestenes84} as said earlier; another good reference is provided by
\citet{Doran03}. The existence of such references allows us to introduce the
vector derivative without further explanation; the reader should search the
cited books for full justification of the definition we give below
\begin{equation}
    \Box = g^\mu \partial_\mu.
\end{equation}
The vector derivative is a vector and can be operated with any multivector
using the established rules; in particular the geometric product of $\Box$ with
a multivector can be decomposed into inner and outer products. When applied to
vector $a$ $(\Box a = \Box \dprod a + \Box \wprod a)$ the inner product is the
divergence of vector $a$ and the outer product is the exterior derivative,
related to the curl although usable in spaces of arbitrary dimension and
expressed as a bivector. We also define the Laplacian as the scalar operator
$\Box^2 = \Box \dprod \Box$. In this work we do not use the conventions of
Riemanian geometry for the affine connection, as was already noted in relation
to Eq.\ (\ref{eq:recframe}). For this reason we will also need to distinguish
between the curved space derivative defined above and the ordinary flat space
derivative
\begin{equation}
    \nabla = \sigma^\mu \partial_\mu = \sum \sigma_\mu \partial_\mu.
\end{equation}
When using spherical coordinates, for instance, the connection will be involved
only in the flat space component of the derivative and we will deal with it by
explicitly expressing the frame vector derivatives.

Velocity is a vector with very special significance in 4DO space because it is
the unitary vector tangent to a geodesic. We therefore attribute high
significance to velocity derivatives, since they express the characteristics of
the particular space we are considering. When the Laplacian is applied to the
velocity vector we obtain a vector
\begin{equation}
    \label{eq:current}
    \Box^2  v = T.
\end{equation}
Vector $T$ is called the \emph{sources vector} and can be expanded into sixteen
terms as
\begin{equation}
    T = {T^\mu}_\nu \sigma_\mu \dot{x}^\nu = (\Box^2 {n^\mu}_\nu)
    \sigma_\mu \dot{x}^\nu.
\end{equation}
The tensor ${T^\mu}_\nu$ contains the coefficients of the sources vector and we
call it the \emph{sources tensor}; it is very similar to the stress tensor of
GR, although its relation to geometry is different. The sources tensor
influences the shape of geodesics but we shall not examine here how such
influence arises, except for very special cases.

Before we begin searching solutions for Eq.\ (\ref{eq:current}) we will show
that this equation can be decomposed into a set of equations similar to
Maxwell's. Consider first the velocity derivative $\Box v = \Box \dprod v +
\Box \wprod v$; the result is a multivector with scalar and bivector part $G =
\Box v$. Now derive again $\Box G = \Box \dprod G + \Box \wprod G$; we know
that the exterior derivative of $G$ vanishes and the divergence equals the
sources vector. Maxwell's equations can be written in a similar form, as was
shown in \citet{Almeida04:2}, with velocity replaced by the vector potential
and multivector $G$ replaced by the Faraday bivector $F$; \citet{Doran03} offer
similar formulation for spacetime.

An isotropic space must be characterised by orthogonal refractive index vectors
$g_\mu$ whose norm can change with coordinates but is the same for all vectors.
We usually relax this condition by accepting that the three $g_j$ must have
equal norm but $g_4$ can be different. The reason for this relaxed isotropy is
found in the parallel usually made with physics by assigning dimensions $1$ to
$3$ to physical space. Isotropy in a physical sense need only be concerned with
these dimensions and ignores what happens with dimension 4. We will therefore
characterise an isotropic space by the refractive index frame $g_j = n_r
\sigma_j$, $g_4 = n_4 \sigma_4$. Indeed we could also accept a non-orthogonal
$g_4$ within the relaxed isotropy concept but we will not do so in this work.

We will only investigate spherically symmetric solutions independent of $x^4$;
this means that the refractive index can be expressed as functions of $r$ in
spherical coordinates. The vector derivative in spherical coordinates is of
course
\begin{equation}
    \Box = \frac{1}{n_r}\, \left(\sigma_r \partial_r + \frac{1}{r}\,
    \sigma_\theta \partial_\theta + \frac{1}{r \sin \theta}\, \sigma_\varphi
    \partial_\varphi \right) + \frac{1}{n_4}\, \sigma_4 \partial_4.
\end{equation}
The Laplacian is the inner product of $\Box$ with itself but the frame
derivatives must be considered
\begin{eqnarray}
    \partial_r \sigma_r = 0, ~~~~ &\partial_\theta \sigma_r = \sigma_\theta,
     ~~~~ &\partial_\varphi \sigma_r = \sin \theta \sigma_\varphi, \nonumber \\
    \partial_r \sigma_\theta = 0, ~~~~ &\partial_\theta \sigma_\theta =
    -\sigma_r, ~~~~ &\partial_\varphi \sigma_\theta = \cos \theta \sigma_\varphi, \\
    \partial_r \sigma_\varphi = 0, ~~~~ &\partial_\theta \sigma_\varphi = 0, ~~~~
    &\partial_\varphi \sigma_\varphi = -\sin \theta\, \sigma_r - \cos \theta\, \sigma_\theta.
    \nonumber
\end{eqnarray}
After evaluation the Laplacian becomes
\begin{equation}
    \label{eq:laplacradial}
    \Box^2 = \frac{1}{(n_r)^2}\, \left(\partial_{rr} + \frac{2}{r}\, \partial_r -
    \frac{n'_r}{n_r}\, \partial_r + \frac{1}{r^2}\, \partial_{\theta \theta}
    +\frac{\cot \theta}{r^2}\, \partial_\theta
    + \frac{\csc^2 \theta}{r^2}\, \partial_{\varphi \varphi} \right) +
    \frac{1}{(n_4)^2}\, \partial_{\tau \tau}.
\end{equation}

In the absence of sources we want the sources tensor to vanish, implying that
the Laplacian of both $n_r$ and $n_4$ must be zero; considering that they are
functions of $r$ we get the following equation for $n_r$
\begin{equation}
    n^{''}_r + \frac{2 n'_r}{r} - \frac{(n'_r)^2}{n_r} = 0,
\end{equation}
with general solution $n_r = b \exp(a/r)$. We can make $b =1$ because we want
the refractive index to be unity at infinity. Using this solution in Eq.\
(\ref{eq:laplacradial}) the Laplacian becomes
\begin{equation}
    \Box^2 = \mathrm{e}^{-a/r}\left(\mathrm{d}^2 + \frac{2}{r}\, \mathrm{d}
     + \frac{a }{r^2}\, \mathrm{d}\right).
\end{equation}
When applied to $n_4$ and equated to zero we obtain solutions which impose $n_4
= n_r $ and so the space must be truly isotropic and not relaxed isotropic as
we had allowed. The solution we have found for the refractive index components
in isotropic space can correctly model Newton dynamics, which led the author to
adhere to it for some time \cite{Almeida01:4}. However if inserted into Eq.\
(\ref{eq:dtau2gen}) this solution produces a GR metric which is verifiably in
disagreement with observations; consequently it has purely geometric
significance.

The inadequacy of the isotropic solution found above for relativistic
predictions deserves some thought, so that we can search for solutions guided
by the results that are expected to have physical significance. In the physical
world we are never in a situation of zero sources because the shape of space or
the existence of a refractive index must always be tested with a test particle.
A test particle is an abstraction corresponding to a point mass considered so
small as to have no influence on the shape of space. But in reality a test
particle is always a source of refractive index and its influence on the shape
of space may not be negligible in any circumstances. If this is the case the
solutions for vanishing sources vector may have only geometric meaning, with no
connection to physical reality.

The question is then how do we include the test particle in Eq.\
(\ref{eq:current}) in order to find physically meaningful solutions. Here we
will make one \emph{add hoc} proposal without further justification because the
author has not yet completed the work that will provide such justification in
geometric terms. The second member of Eq.\ (\ref{eq:current}) will not be zero
and we will impose the sources vector
\begin{equation}
    \label{eq:statpart}
    J = -\nabla^2 n_4 \sigma_4.
\end{equation}
Equation (\ref{eq:current}) becomes
\begin{equation}
    \label{eq:gravitation}
    \Box^2 v = -\nabla^2 n_4 \sigma_4;
\end{equation}
as a result the equation for $n_r$ remains unchanged but the equation for $n_4$
becomes
\begin{equation}
    n^{''}_4 + \frac{2 n'_4}{r} - \frac{n'_r n'_4}{n_r}
    = - n^{''}_4 + \frac{2 n'_4}{r}.
\end{equation}
When $n_r$ is given the exponential form found above the solution is $n_4 =
\sqrt{n_r}$. This can now be entered into Eq.\ (\ref{eq:dtau2gen}) and the
coefficients can be expanded in series and compared to Schwarzschild's for the
determination of parameter $a$. The final solution, for a stationary mass $M$
is
\begin{equation}
    \label{eq:refind}
    n_r = \mathrm{e}^{2M/r},~~~~n_4 = \mathrm{e}^{M/r}.
\end{equation}

Equation (\ref{eq:gravitation}) can be interpreted in physical terms as
containing the essence of gravitation. When solved for spherically symmetric
solutions, as we have done, the first member provides the definition of a
stationary gravitational mass as the factor $M$ appearing in the exponent and
the second member defines inertial mass as $\nabla^2 n_4$. Gravitational mass
is defined with recourse to some particle which undergoes its influence and is
animated with velocity $v$ and inertial mass cannot be defined without some
field $n_4$ acting upon it. Complete investigation of the sources tensor
elements and their relation to physical quantities is not yet done. It is
believed that the 16 terms of this tensor have strong links with homologous
elements of stress tensor in GR but this will have to be verified.

Finally we turn our attention to hyperspherical coordinates. The position
vector is quite simply $x = \tau \sigma_\tau$, where the coordinate is the
distance to the hypersphere centre. Differentiating the position vector we
obtain the displacement vector, which is a natural generalisation of 3D
spherical coordinates case
\begin{equation}
    \mathrm{d}x = \sigma_\tau \mathrm{d} \tau + \tau \sigma_\rho \mathrm{d}
    \rho + \tau \sin \rho \sigma_\theta \mathrm{d}\theta + \tau \sin \rho
    \sin \theta \sigma_\varphi \mathrm{d} \varphi;
\end{equation}
$\rho$, $\theta$ and $\varphi$ are angles. The velocity in an isotropic medium
should now be written as
\begin{equation}
    \label{eq:hypervelocity}
    v = n_4 \sigma_\tau \dot{\tau} + n_r \tau (\sigma_\rho
    \dot{\rho} + \sin \rho \sigma_\theta \dot{\theta} + \sin \rho
    \sin \theta \sigma_\varphi \dot{\varphi}).
\end{equation}

In order to replace the angular coordinate $\rho$ with a distance coordinate
$r$ we can make $r = \tau \rho$ and derive with respect to time
\begin{equation}
    \dot{r} =\rho \dot{\tau} + \tau \dot{\rho}
    = \frac{r}{\tau}\, \dot{\tau} + \tau \dot{\rho}.
\end{equation}
Taking $\tau \dot{\rho}$ from this equation and inserting into Eq.\
(\ref{eq:hypervelocity}), assuming that $\sin \rho$ is sufficiently small to be
replaced by $\rho$
\begin{equation}
    \label{eq:veluniverse}
    v = n_4 \left(\sigma_\tau - \frac{r}{\tau}\, \sigma_r \right) \dot{\tau} +
    n_r (\sigma_r \dot{r} + r \sigma_\theta \dot{\theta}
    + r \sin \theta \sigma_\varphi \dot{\varphi}).
\end{equation}
we have also replaced $\sigma_\rho$ by $\sigma_r$ for consistency with the new
coordinates.

We have just defined a particularly important set of coordinates, which appears
to be especially well adapted to describe the physical Universe, with $\tau$
being interpreted as the Universe's age or its radius; note that time and
distance cannot be distinguished in non-dimensional units. When $r
\dot{\tau}/\tau$ is small in Eq.\ (\ref{eq:veluniverse}), the refractive index
vectors become orthogonal and we use $n_4$ and $n_r$ in conjunction with Eq.\
(\ref{eq:dtau2gen}) to obtain a GR metric whose coefficients are equivalent so
Schwarzschild's on the first terms of their series expansions. When $r
\dot{\tau}/\tau$ cannot be neglected, however, the equation can explain the
Universe's expansion and flat rotation curves in galaxies without dark matter
intervention. A more complete discussion of this subject can be found in Ref.\
\cite{Almeida04:1}.
\section{Conclusions}
Euclidean and Minkowskian 4-spaces can be formally linked through the null
subspace of 5-dimensional space with signature $(-++++)$. The extension of such
formalism to non-flat spaces allows the transition between spaces with both
signatures and the paper discusses some conditions for metric and geodesic
translation. For its similarities with optics, the geometry of 4-spaces with
Euclidean signature is called 4-dimensional optics (4DO). Using only geometric
arguments it is possible to define such concepts as velocity and trajectory in
4DO which become physical concepts when proper and natural assignments are
made.

One important point which is addressed for the first time in the author's work
is the link between the shape of space and the sources of curvature. This is
done on geometrical grounds but it is also placed in the context of physics.
The equation pertaining to the test of gravity by a test particle is proposed
and solved for the spherically symmetric case providing a solution equivalent
to Schwarzschild's as first approximation. Some mention is made of
hyperspherical coordinates and the reader is referred to previous work linking
this geometry to the Universe's expansion in the absence of dark matter.
  \bibliographystyle{unsrtbda}
  \bibliography{Abrev,aberrations,assistentes}   

\begin{thebibliography}{11}
\expandafter\ifx\csname natexlab\endcsname\relax\def\natexlab#1{#1}\fi
\expandafter\ifx\csname bibnamefont\endcsname\relax
  \def\bibnamefont#1{#1}\fi
\expandafter\ifx\csname bibfnamefont\endcsname\relax
  \def\bibfnamefont#1{#1}\fi
\expandafter\ifx\csname citenamefont\endcsname\relax
  \def\citenamefont#1{#1}\fi
\expandafter\ifx\csname url\endcsname\relax
  \def\url#1{\texttt{#1}}\fi
\expandafter\ifx\csname urlprefix\endcsname\relax\def\urlprefix{URL }\fi
\providecommand{\bibinfo}[2]{#2}
\providecommand{\eprint}[1]{\href{http://www.arxiv.org/abs/#1}{\texttt{#1}}}

\bibitem[{\citenamefont{Hestenes and Sobczyk}(1989)}]{Hestenes84}
\bibinfo{author}{\bibfnamefont{D.}~\bibnamefont{Hestenes}} \bibnamefont{and}
  \bibinfo{author}{\bibfnamefont{G.}~\bibnamefont{Sobczyk}},
  \emph{\bibinfo{title}{Clifford Algebras to Geometric Calculus. A Unified
  Language for Mathematics and Physics}}, Fundamental Theories of Physics
  (\bibinfo{publisher}{Reidel}, \bibinfo{address}{Dordrecht},
  \bibinfo{year}{1989}).

\bibitem[{\citenamefont{Gull et~al.}(1993)\citenamefont{Gull, Lasenby, and
  Doran}}]{Gull93}
\bibinfo{author}{\bibfnamefont{S.}~\bibnamefont{Gull}},
  \bibinfo{author}{\bibfnamefont{A.}~\bibnamefont{Lasenby}}, \bibnamefont{and}
  \bibinfo{author}{\bibfnamefont{C.}~\bibnamefont{Doran}},
  \emph{\bibinfo{title}{Imaginary numbers are not real. --- {T}he geometric
  algebra of spacetime}}, \bibinfo{journal}{Found. Phys.}
  \textbf{\bibinfo{volume}{23}}, \bibinfo{pages}{1175}, 1993,
  \urlprefix\url{http://www.mrao.cam.ac.uk/~clifford/publications/abstracts/im%
ag\_numbs.html}.

\bibitem[{\citenamefont{Lasenby and Doran}(1999)}]{Lasenby99}
\bibinfo{author}{\bibfnamefont{A.}~\bibnamefont{Lasenby}} \bibnamefont{and}
  \bibinfo{author}{\bibfnamefont{C.}~\bibnamefont{Doran}},
  \emph{\bibinfo{title}{Physical applications of geometric algebra}},
  \bibinfo{howpublished}{handout collection from a {C}ambridge {U}niversity
  lecture course}, 1999,
  \urlprefix\url{http://www.mrao.cam.ac.uk/~clifford/ptIIIcourse/course99/}.

\bibitem[{\citenamefont{Doran and Lasenby}(2003)}]{Doran03}
\bibinfo{author}{\bibfnamefont{C.}~\bibnamefont{Doran}} \bibnamefont{and}
  \bibinfo{author}{\bibfnamefont{A.}~\bibnamefont{Lasenby}},
  \emph{\bibinfo{title}{Geometric Algebra for Physicists}}
  (\bibinfo{publisher}{Cambridge University Press},
  \bibinfo{address}{Cambridge, U.K.}, \bibinfo{year}{2003}).

\bibitem[{\citenamefont{Hestenes}(2003)}]{Hestenes03}
\bibinfo{author}{\bibfnamefont{D.}~\bibnamefont{Hestenes}},
  \emph{\bibinfo{title}{New Foundations for Classical Mechanics}}
  (\bibinfo{publisher}{Kluwer Academic Publishers},
  \bibinfo{address}{Dordrecht, The Netherlands}, \bibinfo{year}{2003}),
  \bibinfo{edition}{2nd} ed.

\bibitem[{\citenamefont{Martin}(1988)}]{Martin88}
\bibinfo{author}{\bibfnamefont{J.~L.} \bibnamefont{Martin}},
  \emph{\bibinfo{title}{General Relativity: A Guide to its Consequences for
  Gravity and Cosmology}} (\bibinfo{publisher}{Ellis Horwood Ltd.},
  \bibinfo{address}{U. K.}, \bibinfo{year}{1988}).

\bibitem[{\citenamefont{Almeida}(2002)}]{Almeida02:2}
\bibinfo{author}{\bibfnamefont{J.~B.} \bibnamefont{Almeida}},
  \emph{\bibinfo{title}{K-calculus in 4-dimensional optics}}, 2002,
  \eprint{physics/0201002}.

\bibitem[{\citenamefont{Montanus}(2001)}]{Montanus01}
\bibinfo{author}{\bibfnamefont{J.~M.~C.} \bibnamefont{Montanus}},
  \emph{\bibinfo{title}{Proper-time formulation of relativistic dynamics}},
  \bibinfo{journal}{Found. Phys.} \textbf{\bibinfo{volume}{31}},
  \bibinfo{pages}{1357}, 2001.

\bibitem[{\citenamefont{Almeida}(2004{\natexlab{a}})}]{Almeida04:1}
\bibinfo{author}{\bibfnamefont{J.~B.} \bibnamefont{Almeida}},
  \emph{\bibinfo{title}{An hypersphere model of the universe -- {T}he dismissal
  of dark matter}}, 2004, \eprint{physics/0402075}.

\bibitem[{\citenamefont{Almeida}(2004{\natexlab{b}})}]{Almeida04:2}
\bibinfo{author}{\bibfnamefont{J.~B.} \bibnamefont{Almeida}},
  \emph{\bibinfo{title}{Maxwell's equations in 4-dimensional {E}uclidean
  space}}, 2004, \eprint{physics/0403058}.

\bibitem[{\citenamefont{Almeida}(2001)}]{Almeida01:4}
\bibinfo{author}{\bibfnamefont{J.~B.} \bibnamefont{Almeida}},
  \emph{\bibinfo{title}{4-dimensional optics, an alternative to relativity}},
  2001, \eprint{gr-qc/0107083}.

\end{thebibliography}
\end{document}